\documentstyle[12pt,aaspp4,tighten]{article}

\def\fun#1#2{\lower3.6pt\vbox{\baselineskip0pt\lineskip.9pt
  \ialign{$\mathsurround=0pt#1\hfil##\hfil$\crcr#2\crcr\sim\crcr}}}

\def\mass{{\cal M}}

\def\Msolar{{\mass_\odot}}
\def\Ampl{{\cal A}}
\def\thresh{{\rm thresh}}
\def\max{{\rm max}}
\def\min{{\rm min}}
\def\inf{{\rm inf}}

\slugcomment{\vbox{\hbox{CU-TP-744}
             \hbox{CAL-603}
             \hbox{astro-ph/9604144}}}

\begin{document}
\title{Rates for Parallax-Shifted Microlensing Events from Ground-Based
Observations of the Galactic Bulge}

\author{Ari Buchalter$^*$\altaffilmark{1} and Marc
Kamionkowski$^\dagger$\altaffilmark{2}}
\affil{$^*$Department of Astronomy, Columbia University, New
York, NY 10027}
\affil{$^\dagger$Department of Physics, Columbia University, New
York, NY 10027}
\altaffiltext{1}{ari@parsifal.phys.columbia.edu}
\altaffiltext{2}{kamion@phys.columbia.edu}

\begin{abstract}

The parallax effect in ground-based microlensing (ML) observations consists of
a distortion to the standard ML light curve arising from the Earth's orbital
motion. This can be used to partially remove the degeneracy among the system
parameters in the event timescale, $t_0$. In most cases, the resolution in
current ML surveys is not accurate enough to observe this effect, but
parallax could conceivably be detected with frequent followup observations of
ML events in progress, providing the photometric errors are small
enough. We calculate the expected fraction of ML events where the
shape distortions will be observable by such followup observations, adopting
Galactic models for the lens and source distributions which are consistent with
observed microlensing timescale distributions. We study the dependence of the
rates for parallax-shifted events on the frequency of followup observations and
on the precision of the photometry. For example, we find that for hourly
observations with typical
photometric errors of 0.01 mag, 6\% of events where the lens is in the
bulge, and 31\% of events where the lens is in the disk, (or $\approx 10$\% of
events overall) will give rise to a measurable parallax shift at the 95\%
confidence level. These fractions may be increased by improved
photometric accuracy and increased sampling frequency. While long-duration
events are favored, the surveys would be effective in picking out such
distortions in events with timescales as low as $t_0 \approx 20$ days. We study
the dependence of these fractions on the assumed disk mass function, and find
that a higher parallax incidence is favored by mass functions with higher mean
masses. Parallax measurements yield the reduced transverse speed, $\tilde{v}$,
which gives both the relative transverse speed and lens mass as a function of
distance. We give examples of the accuracies with which $\tilde{v}$ may be
measured in typical parallax events. Fitting ML light curves which may be
shape-distorted (e.g., by parallax, blending, etc.) with only the 3 standard
ML parameters can result in
inferred values for these quantities which are significantly in error. Using
our model, we study the effects of such systematic errors and find that, due
primarily to blending, the inferred timescales from such fits, for events with
disk lenses, tend to shift the event duration distribution by $\approx 10$\%
towards shorter $t_0$. Events where the lens resides in the bulge are
essentially unaffected. In both cases, the impact-parameter distribution is
depressed slightly at both the low and high ends.

\end{abstract}



\section{INTRODUCTION}

The recent observations by the MACHO (Alcock et al.~1995a, Alcock et al.~1996),
EROS (Aubourg et al.~1993), OGLE (Udalski et al.~1994a), and DUO (Alard et
al.~1995) collaborations of gravitational microlensing (ML) of stars in the LMC
and Galactic bulge have generated tremendous excitement in astrophysics. These
surveys provide a new probe of Galactic structure and low-mass stellar
populations. ML observations of the LMC allow measurements of the dark-matter
content of the Galactic halo (Bennett et al.~1996), placing important
constraints on dark-matter theories. Also intriguing have been observations of
ML events towards the Galactic center, which probe the inner disk and bulge of
our Galaxy. The roughly 100 events observed to date imply an optical depth
towards the bulge of ($3.3 \pm 1.2$) $\times 10^{-6}$, three times that
predicted by theoretical estimates (Griest et al.~1991), and reveal the need
for a better understanding of Galactic structure.

One explanation for the observed excess of bulge events would be the presence
of  a hitherto undiscovered population of compact, sub-stellar objects,
implying the existence of more mass in the disk than previously believed
(Alcock et al.~1994; Gould, Miralda-Escude, \& Bahcall 1994). This would
require an upturn in the stellar mass function (MF) below the hydrogen-burning
limit. Another possibility is that the lenses comprise an ordinary stellar
population, and the enhancement of ML events is due to non-axisymmetric
structure, such as a bar, in the Galactic bulge. The optical depth calculated
{}from a self-consistent bar model, matching both kinematic observations of the
bulge and the COBE image of the Galaxy, is consistent with the MACHO and OGLE
results (Kiraga \& Paczy\'{n}ski 1994; Zhao, Spergel, \& Rich 1995; Zhou, Rich,
\& Spergel 1995). Furthermore, the best-fit models to the observed ML duration
distribution have a median lens mass of about $0.2\ \Msolar$, {\em above} the
hydrogen-burning limit, and consistent with ordinary stellar populations.

Determining the nature and location of the lenses will enable us to learn
whether the enhancement of bulge events is due to lenses in the bar or disk,
and probe a region of the stellar MF about which little is known. However,
while numerous cases of ML have been observed, essential information such as
the mass, distance, and velocity distributions of the lensing population can
only be disentangled in a statistical manner, due to the degeneracy of
microlensing light curves with respect to these parameters. For an ideal ML
event, where the source and lens are assumed to be point-like,
the lens is assumed to be dark, and the velocities of the observer, source, and
lens are constant, the amplification is given by
\begin{equation}
     A[u(t)] = \frac {u^2 +2} {u(u^2 +4)^{1/2}};\qquad  u(t)=
     \left[  \left( \frac {t-t_{\max}} {t_0} \right)^2 +
     u_{\min}^2\right]^{1/2},
\label{standardamplification}
\end{equation}
where $u$ is the impact parameter from the observer-source line to the lens in
units of the Einstein radius, ${t_0} = R_{e}/v$ is the event timescale, $v$ is
the transverse speed of the lens relative to the source-star line of
sight, and $t_{\max}$ is the time at which peak amplification,
$A_\max = A(u_{\min})$, occurs. The Einstein radius, $R_{e}$, is determined by
the geometry of the event and the lens mass, and is given by
\begin{equation}
     R_e = \left[ {4G\mass_l D_{ol} (D_{os} - D_{ol}) \over c^2 D_{os}}
     \right]^{1/2},
\label{einsteinradius}
\end{equation}
where $\mass_l$ is the lens mass and $D_{ol}$ and $D_{os}$ denote the
observer-lens and observer-source distances, respectively. We see from
Eq.~(\ref{standardamplification}) that the ML light curve can be fit by the
three parameters $t_0$, $t_{\max}$, and $u_{\min}$. The latter two, however,
only tell when the lens passed nearest to the line of sight and how close it
came, revealing little useful information about the event. All the parameters
of interest, namely  $\mass_l$, $v$, $D_{ol}$, and $D_{os}$, are folded into
the single parameter $t_0$. (In practice, $D_{os}$ may be obtained to
reasonable accuracy.)

Several techniques for breaking the degeneracy have been proposed, many of
which involve fitting for distortions to the generic ML light curve which may
be detected in some fraction of observed events (Nemiroff \& Wickramasinghe
1994; Witt \& Mao 1994; Maoz \& Gould 1994; Loeb \& Sasselov 1995; Han \& Gould
1995; Gaudi \& Gould 1996). These distortions arise from violations of the
standard ML assumptions. For example, if the lenses consist partly of ordinary,
low-mass dwarfs in the Galactic disk and bulge, the contribution of the lens
light would distort the observed light curve differently in different
wavebands, violating the assumption of achromaticity, and distorting the shape,
[c.f. Eq.~(\ref{standardamplification})] of the light curve (Kamionkowski 1995;
Buchalter, Kamionkowski, \& Rich 1995, hereafter BKR). Ground-based
observations of this color-shift effect in two or more wavebands can remove
entirely the degeneracy in $t_0$. If the lenses are not completely dark, then
color shifting should, in principle, affect every event and the observed
incidence of this effect is simply a
function of resolution of the data (namely sampling frequency and level of
photometric error). The MACHO
collaboration has already observed another type of distortion, where the time
symmetry of the light curve is broken by the Earth's orbital motion (Alcock et
al.~1995b). This parallax effect allows one to compare the projected Einstein
ring with the size of the Earth's orbit and thereby obtain an additional
constraint relating $\mass_l$, $v$, and $D_{ol}$, effectively removing one
degree of degeneracy. Strictly speaking, this effect is also present in every
event, but is not expected to be frequently observed by any of the existing ML
surveys; the light curves are sampled too infrequently and the photometric
errors are too large. The single confirmed parallax event detected to date was
fortuitous in that it was both long enough ($t_0 \approx 110$ days) for the
light curve to be heavily sampled, and well-situated during the MACHO observing
season so that the asymmetry could be fit along the entire light curve. In
addition, the effect is more dramatic for longer events, during which the Earth
can move through an appreciable fraction of its orbit. It is only for these
rarer long events
that a parallax shift may be observed by the low-resolution ML surveys.
However, with the early-warning alert systems developed by both the MACHO and
OGLE groups (Stubbs et al. 1994; Udalski et al. 1994b), it is conceivable that
a program of followup observations with frequent and precise measurements could
measure the light curves with sufficient accuracy to detect the parallax effect
in a larger
fraction of events. The PLANET (Albrow et al.~1995) and GMAN (Pratt et
al.~1996) collaborations are currently performing such observations, with the
primary purpose of detecting planets around lenses, and Tytler et al. (1996)
are proposing another such search. Planetary masses would give rise to smaller
event timescales and thus produce narrow spikes on the lensing light curve
which likely go undetected in current surveys. These spikes could be resolved
with observations by dedicated telescopes performing rapid sampling of events
in progress with high
photometric precision. Such ground-based surveys are well-suited to pick out
the parallax effect for shorter-term events, as well as distortions due to
unlensed light from the lens.

In this paper, we determine the fraction of ML events toward the Galactic bulge
which should show a measurable parallax shift in such followup programs. We
calculate the fractions which will arise if the lenses are all in the bulge and
if the lenses are all in the disk, using realistic models for the lens
distributions which are consistent with the currently observed ML timescale
distribution. A Monte Carlo technique is used to simulate events and generate
parallax-distorted light curves for each. It is then determined for each event
whether the distorted light curve can be distinguished at the 95\% confidence
level from a standard ML light curve. The calculation is performed for several
values of the sampling frequency and for several values of photometric
accuracies that may be attainable. Our results indicate that, depending on the
survey sensitivity, the expected fraction of parallax-shifted events
($F_{PSE}$) ranges from 0.06 to 0.31 for bulge self-lensing events, 0.31 to
0.77 for disk-lensing-bulge events, and 0.49 to 0.83 for (rare) disk
self-lensing events, with current-generation followup surveys favoring the
lower values. Since the parallax effect is particularly important for the case
of disk lenses, we also examine the dependence of this effect on the adopted
mass function (MF) for the disk, and find higher incidences from MFs with
higher mean masses. We further calculate the dependence of $F_{PSE}$ on event
timescale, finding that the intensive followup programs should measure a
substantial contribution to $F_{PSE}$ from events with $t_0$ as low as 20 days,
and we also examine what may be learned from a typical PSE. We emphasize that,
like color-shift analysis, parallax analysis can be directly applied to all
events, so that information about the mass, velocity and spatial distributions
of the lenses becomes available on an event by event---rather than on a
statistical---basis.

Although parallax and blending effects may be present to some degree in a large
fraction of events, they are expected to be small in most cases, so that a
standard three-parameter fit to a slightly distorted light curve may be deemed
adequate. However, fitting the standard ML amplification function to a light
curve distorted by these effects can result in systematic errors in the
inferred fit parameters, most notably $t_0$ and $u_{\min}$. These inaccuracies
may lead to a systematic miscalculation of the duration and amplification
distributions, and of the overall optical depth. Thus, we perform another
simulation to generate shape-distorted light curves including both parallax and
color-shift effects (i.e., blending due to the lens). A standard
three-parameter fit to these light curves is then applied, and the inferred
parameters are used to compare the resulting amplification and event duration
distributions with their actual values. We find for the latter that the errors
incurred by three-parameter fits are negligible for bulge lenses, but may be
quite large ($\geq 10\%$) if the lenses are in the disk, particularly if the
disk MF declines at the low-mass end. We also find that errors in the inferred
minimum impact parameter lead to a non-uniform distribution in peak
amplification.

The plan of the paper is as follows: In Section 2, we review the
characteristics of parallax-distorted events. In Section 3, we summarize the
models and techniques used in the calculation and present the results of our
calculations of expected fraction of PSEs. In Section 4, we examine the impact
of parameterizing shape-distorted light curves with the standard ML fit on the
inferred ML parameters, and in Section 5 we make some concluding remarks.

\section{PARALLAX-SHIFTED MICROLENSING EVENTS}

To parameterize the parallax effect, consider first a standard ML event as
viewed at the position of the Sun. In that case, we regain the
constant-velocity condition, so that the scaled distance from the lens to the
line of sight, $u(t)$, is properly described by
Eq.~(\ref{standardamplification}), where $u_{\min}$ now represents the minimum
distance from the lens to the Sun-source line of sight. We can modify the
expression for $u(t)$ to incorporate the Earth's orbital motion by projecting
to the position of the lens the motion of the observer (Earth) along the
ecliptic. Figure 1 depicts a simplified version
of the geometry where we have assumed, for the sake of illustration, that the
ecliptic is perpendicular to the Sun-source line. In this case, we see from
Figure 1(a)  that the projection of the Earth's orbit traces out a circle with
a radius (scaled in units of $R_e$) given by
\begin{equation}
\alpha = \frac{a}{R_e} = \frac {R_{\oplus}} {R_e} (1-x),
\label{alpha}
\end{equation}
\begin{figure}
\plotone{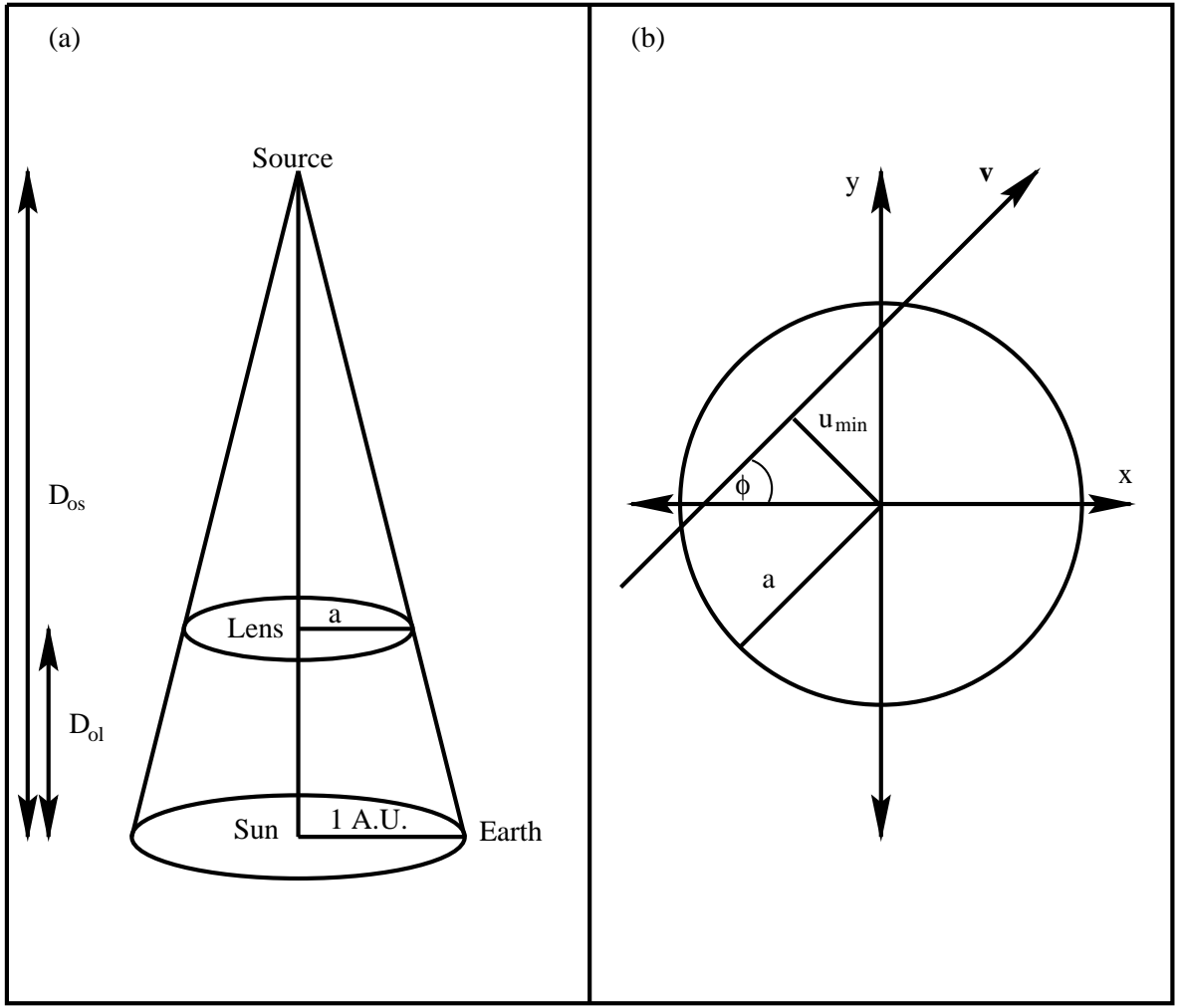}
\caption{(a) Simplified geometry of the parallax cone, assuming the ecliptic is
perpendicular to the Sun-source line. The Earth's orbit, projected to the
position of the lens, traces out a circle of radius $a$. (b) Cross section of
the parallax cone at the distance of the lens. The lens velocity is denoted by
${\bf v}$. Note that $u_{\min}$ is no longer the minimum distance from the lens
to the observer-source line of sight, but rather to the Sun-source line of
sight.}
\end{figure}
where $R_{\oplus} = 1$ A.U. and $x = D_{ol}/D_{os}$. Figure 1(b) shows a cross
section of the parallax cone in Figure 1(a), taken at the position of the lens.
The projected motion of the Earth at the lens position is given by
${\bf r_{\oplus}}(t) =
\{\alpha\cos[\Omega(t-t_c)],\alpha\sin[\Omega(t-t_c)]\}$. In actuality, we must
account for the true orientation of the ecliptic with respect to the Galactic
plane. The true projection of the Earth's orbit is an ellipse whose
eccentricity is determined by the ecliptic latitude, $\beta$, of the source, so
that
\begin{equation}
{\bf r_{\oplus}}(t) = \{\alpha\sin{\beta}\cos{[\Omega(t-t_c)]},\;
\alpha\cos{\beta}\sin{[\Omega(t-t_c)]}\}
\label{rearth}
\end{equation}
is the correct expression. For stars in Baade's Window (${\ell} = -1^{\circ},
b=4^{\circ}$), $\beta \approx 4^{\circ}$, so $\cos{\beta} \approx 1$.
The parameter $\Omega$ is given by
\begin{equation}
\Omega(t-t_c) = \Omega_{0}(t-t_c) + 2\epsilon\sin[\Omega_{0}(t-t_p)],
\label{Omega}
\end{equation}
where $t_c$ is the time when the Earth is closest to the Sun-source line,
$\Omega_{0} = 2\pi$ yr$^{-1}$, $t_p$ is the time of perihelion and $\epsilon =
0.017$ is the eccentricity of Earth's orbit. We include the dependence on
$\epsilon$ in our expression for $\Omega$ since it may contribute appreciably
near $t=t_c$, though we omit it in the expression for $\alpha$, where it is
always negligible. The position of the lens is given by Figure 1(b) as
\begin{equation}
{\bf{r}_{\rm lens}}(t) = \{-u_{\min}\sin{\phi} + v(t-t_{\max})\cos{\phi},\;
u_{\min}\cos{\phi} + v(t-t_{\max})\sin{\phi}\},
\label{rlens}
\end{equation}
where $\phi$ denotes the angle between ${\bf v}$ and ecliptic north. Using
$\omega = t_{0}^{-1}$, we thus have
\begin{eqnarray}
u^{2}(t) & = & (x_{\rm lens} - x_{\oplus})^2 + (y_{\rm lens} - y_{\oplus})^2
\nonumber \\
         & = & u_{\min}^2 + {\omega}^{2}(t-t_{\max})^2 \nonumber \\
         &   & + {\alpha}^{2}\sin^{2}{[\Omega(t-t_c)]} +
{\alpha}^{2}\sin^{2}{\beta}\cos^{2}{[\Omega(t-t_c)]} \\
         &   & - 2{\alpha}\sin{[\Omega(t-t_c)]}[\omega(t-t_{\max})\sin{\phi}
+ u_{\min}\cos{\phi}] \nonumber \\
         &   & + 2{\alpha}\sin{\beta}\cos{[\Omega(t-t_c)]}[u_{\min}\sin{\phi}
- \omega(t-t_{\max})\cos{\phi}] \nonumber
\label{u(t)}
\end{eqnarray}
so that a parallax-shifted light curve is parameterized by the 5 quantities
$t_{\max}$, $t_0$, $u_{\min}$, $\alpha$, and $\phi$. In our simulations, we
include an additional parameter to account for the presence of unlensed light
(either due to the lens or a chance blended star) in the light curve.

Figure 2 illustrates an example of a ML light curve distorted by the parallax
effect. The solid curve is a simulated parallax-shifted event (PSE), arising
{}from a bulge self-lensing configuration with $t_0 = 100$ days and $\alpha =
0.13$. The dashed curve is a standard ML curve with the same peak height
($A_{\max}$), peak time ($t_{\max}$), and FWHM, and has a timescale of $t_0 =
97$ days. It is important to note that since the two curves have different
shapes, as indicated by the residuals in the lower panel, a fit to a PSE which
involves only the parameters $t_{\max}$, $t_0$, and $u_{\min}$ can generate
inferred values of these quantities which are significantly in error. This is
further evidenced by the MACHO collaboration analysis of their PSE, in which
they find the best-fit standard curve to have $t_0 = 141$ days and $u_{\min} =
0.101$, and the best-fit parallax curve to have $t_0 = 111$ days and
$u_{\min} = 0.159$ (Alcock et al.~1995b). Failing to account for distortions in
observed ML light curves can lead to errors in the inferred timescales and thus
in the inferred optical depth as well.
\begin{figure}
\plotone{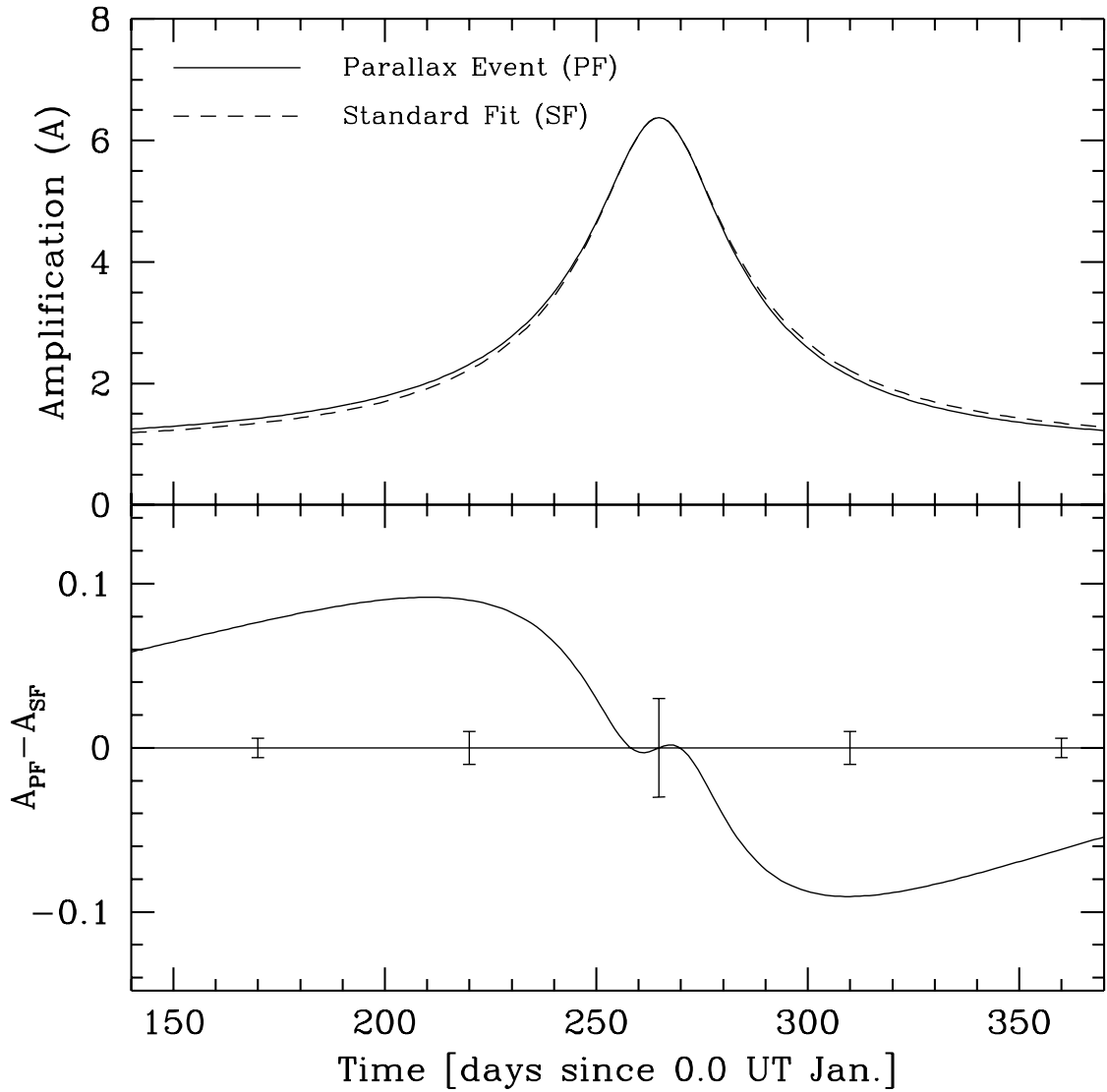}
\caption{The upper panel shows a parallax-shifted light curve (solid line) for
an event arising from a 0.3-$\Msolar$ object at 6 kpc (i.e., the near end of
the bulge) lensing a source at 8 kpc. This curves has the values $\alpha =
0.13$ and $t_0 = 100$ days. A standard ML light curve (dashed line), with
identical $A_{\max}$, $t_{\max}$, and FWHM, has a value of $t_{\max} = 97$
days. The residuals in the lower panel illustrate the time asymmetry of the
parallax effect in this event.}
\end{figure}

\section{EXPECTED FRACTIONS OF PARALLAX-SHIFTED EVENTS}

To calculate the expected fraction of ML events towards Baade's window which
will give rise to a measurable parallax shift, we employ the method used by BKR
in a similar calculation of expected rates for color-shifted events. The
Galactic model consists of a rapidly-rotating bar extending out to 3 kpc (Zhao
1996) and a double exponential disk, extending from 3 kpc out to the solar
circle, with a rotational velocity of 220 km s$^{-1}$ coupled with
position-dependent velocity dispersions of order 30 km s$^{-1}$. Both
components are populated by stars following a logarithmic MF in the range $0.1
{\Msolar} \leq {\mass_l} \leq 1.2 {\Msolar}$. These distributions produce an
event duration distribution which is consistent with observed ML timescales
(see BKR, Figure 3), though other evidence points towards a disk MF which
exhibits some flattening below 1 ${\Msolar}$ (Scalo 1986; Larson 1986; Gould,
Bahcall, \& Flynn 1995).

Using this model, we wish to predict the incidence of PSEs that would be
observed by the followup programs described above. We parameterize various
observing strategies by the time interval between successive measurements and
by the level of photometric error. Since a constant error in magnitude
translates to a constant percent error in amplification, we assume that any
given value of $A$ has a gaussian distribution centered at the true value with
variance $\sigma_{i} \approx f A(t_i)$ which is some fraction $f$ of the
true amplification, as determined by the assumed magnitude error. To determine
whether a shape distortion in a given event will be observable, we first
simulate the expected observed amplification with light-curve measurements for
a given sampling frequency and fractional error, and then ask whether the
6-parameter fit can recover a value of $\alpha$ that is
different from zero at the 95\% ($2{\sigma}$) confidence level. If $\alpha$
differs from zero
with this statistical significance, then a shape distortion from
the parallax effect has been observed. All events were weighted by the
frequency corresponding to their timescale, $t_0$, (see BKR, Section 4) and
summed to give an overall normalization. The fraction of parallax-shifted
events, $F_{PSE}$, is defined as the ratio of the weighted sum of all PSEs to
the overall normalization. In our calculations, we adopt the OGLE detection
efficiency, given by ${\epsilon}(t_0)=0.3\exp[-(t_0/11 \mbox{ days})^{-0.7}]$
(Udalski et al.~1994a). The MACHO efficiencies are typically higher for a given
event duration (Alcock et al.~1996), but do not significantly alter our
results for parallax shifting.

In determining $F_{PSE}$, we consider separately the case of bulge self-lensing
(BB), disk objects lensing bulge sources (DB), and disk self-lensing (DD).
Although the disk lensing events contribute less to the overall optical depth
(particularly in the case of disk self-lensing), the lower transverse speed of
the disk lenses will give rise to longer-duration events whose light curves
will be heavily sampled and thus more likely to yield a parallax shift. Our
light-curve analysis for each event included data from the time, $t_{s}$, at
which the amplification rose above threshold (i.e., $A(t_{s}) = A_T = 1.34$)
until $3t_0$ after the amplification dropped below threshold. Our results
indicate that additional data out to $3t_0$ gives rise to a considerable
increase in $F_{PSE}$ compared to data taken only while $A \geq A_T$, since the
effects of parallax distortion may be significant away from the peak; extending
the data out to $7t_0$, however, does not appreciably increase $F_{PSE}$ beyond
the $3t_0$ value.
Our simulations included only those events in which the event timescale
$t_0$ satisfied $t_0 \leq$ 1 year. For each event, a $u_{\min}$ was chosen
randomly in the range $0 < u_{\min} < 1$. If, however, the experimental
efficiency for detecting low-amplification events is small, so that the
effective range of $u_{\min}$ is, say from $0$ to $0.7$ (Alcock et al.~ 1996;
Gaudi \& Gould 1996), then the fraction of parallax-shifted events would
increase since the effect becomes more pronounced at higher amplifications.

Figure 3 displays the results of our calculation, showing $F_{PSE}$ as a
function of sampling interval, for the various lens/source configurations. The
solid, short-dashed, and long-dashed curves represent photometric errors of
0.05, 0.01, and 0.001 mag, respectively. The results indicate that current
surveys, which sample roughly daily with typical errors of order 0.05 mag, are
not expected to yield many PSEs. For these observational parameters, about 1\%
of all events should exhibit a detectable parallax shift, consistent with the
one PSE observed by the MACHO collaboration out of a total of about 100 bulge
events to date. However, the Figure shows significant increases in $F_{PSE}$
for decreased sampling intervals, and lower levels of photometric error. For
the most likely case of bulge self-lensing (BB), hourly sampling with 0.01-mag
errors, typical of the PLANET experiment, gives $F_{PSE}$ = 0.06, increasing to
0.10 for 15-minute sampling. In the case of DB lensing, these numbers increase
to 0.31 and 0.
42, respectively. Together, these imply an overall expected parallax-shifting
rate of 10\% for hourly sampling and 15\% for 15-minute sampling (assuming
$\tau_{DB} \approx \frac{1}{5}\tau_{BB}$ and neglecting $\tau_{DD}$). Depending
on their capabilities, current planetary surveys are thus expected to detect
parallax effects in roughly 10\% to 15\% of all events.
It is, however, conceivable that future experiments with high-precision
relative photometry from additional dedicated telescopes can produce errors in
the millimagnitude range (Gilliland et al.~1993; Ciardullo \& Bond 1996).
Differencing techniques have already been used to achieve extraordinary
precision in pixel lensing experiments (Tomaney \& Crotts 1996). To
illustrate what may be learned by such experiments, we also consider the
results from millimagnitude errors. Together with a rapid sampling rate, this
improvement could bring the expected $F_{PSE}$ as high as 0.31 for BB events
and 0.77 for DB events, for a maximum overall rate of 39\%.
In general, events arising from disk self-lensing (DD) are expected to be rare,
since $\tau_{DD}$ is only about $5 \times 10^{-3}\tau_{BB}$. Owing to their low
relative transverse speeds however, any DD events observed will have a high
likelihood of exhibiting the parallax effect. Indeed, for 0.01-mag errors,
$F_{PSE} = 0.49$ for hourly sampling, increasing to 0.59 for 15-minute
sampling. It should be noted that if the unlensed-light parameter is neglected
in the fitting procedure, the above values for $F_{PSE}$ in the different
scenarios increase
substantially, doubling in some cases. The unlensed-light parameter in each
waveband introduces another degree of freedom in the fit, which drives down
$F_{PSE}$. Since we have no {\em a priori} knowledge of the extent to which a
given observed ML light curve is distorted by these different effects, it is
important to include the possible contribution of blended light in the fit.
\begin{figure}
\plotone{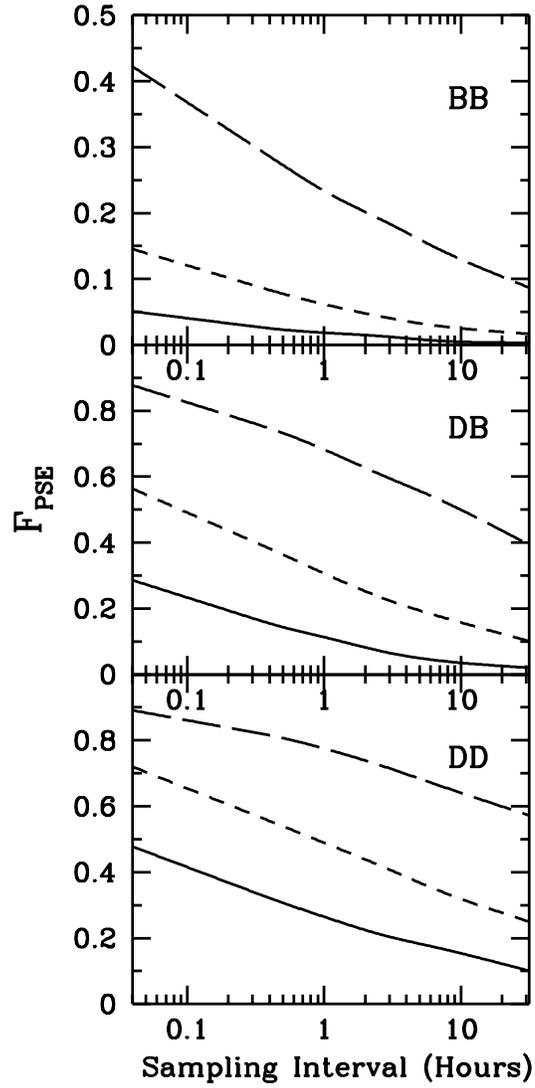}
\caption{Expected fraction of PSEs as a function of sampling interval. The
solid, short-dashed, and long-dashed curves correspond to photometric errors of
0.05, 0.01, and 0.001 mag, respectively. The various configurations are denoted
by BB for bulge self-lensing, DB for disk-lensing-bulge events, and DD for disk
self-lensing.}
\end{figure}

In Figure 4 we plot $F_{PSE}$ as a function of $t_0$ for the three lens/source
configurations, assuming 0.5-hourly sampling with 0.005-mag errors (thin
lines), and daily sampling with 0.05-mag errors (thick lines). It is clear from
the Figure that the intensive followup programs will provide a vast improvement
over the existing large surveys in detecting parallax distortions not only for
the longest duration events, but for all events with timsescales $t_0 \geq 10$
days. Moreover, the Figure shows that the parallax effect is more pronounced
for
events in which the lens resides in the Galactic disk (i.e., DB and DD). This
is due to two factors: the lower transverse speeds arising from our co-rotation
with disk lenses give rise to longer-duration events which can be heavily
sampled along a larger fraction of the Earth's orbit (affecting DB and DD), and
the typical separations between disk stars (lenses) and bulge stars (sources)
give rise to a more favorable parallax geometry (affecting DB only) (see Figure
1). For 0.5-hourly sampling with 0.005-mag errors, $F_{PSE}$ for BB events
becomes appreciable for timescales of $t_0 \geq 20$ days, reaching 0.5 near
$t_0 = 35$ days, while for DB events $F_{PSE}$ becomes appreciable for $t_0
\geq 15$ days, reaching 0.5 at $t_0 = 26$ days. Note that while the overall
$F_{PSE}$ is greater for DD events, as in Figure 3, the fraction as a function
of $t_0$ is lower, especially for smaller $t_0$, because the duration
distribution for DD events is
shifted towards higher $t_0$ owing to their lower relative transverse speeds.
\begin{figure}
\plotone{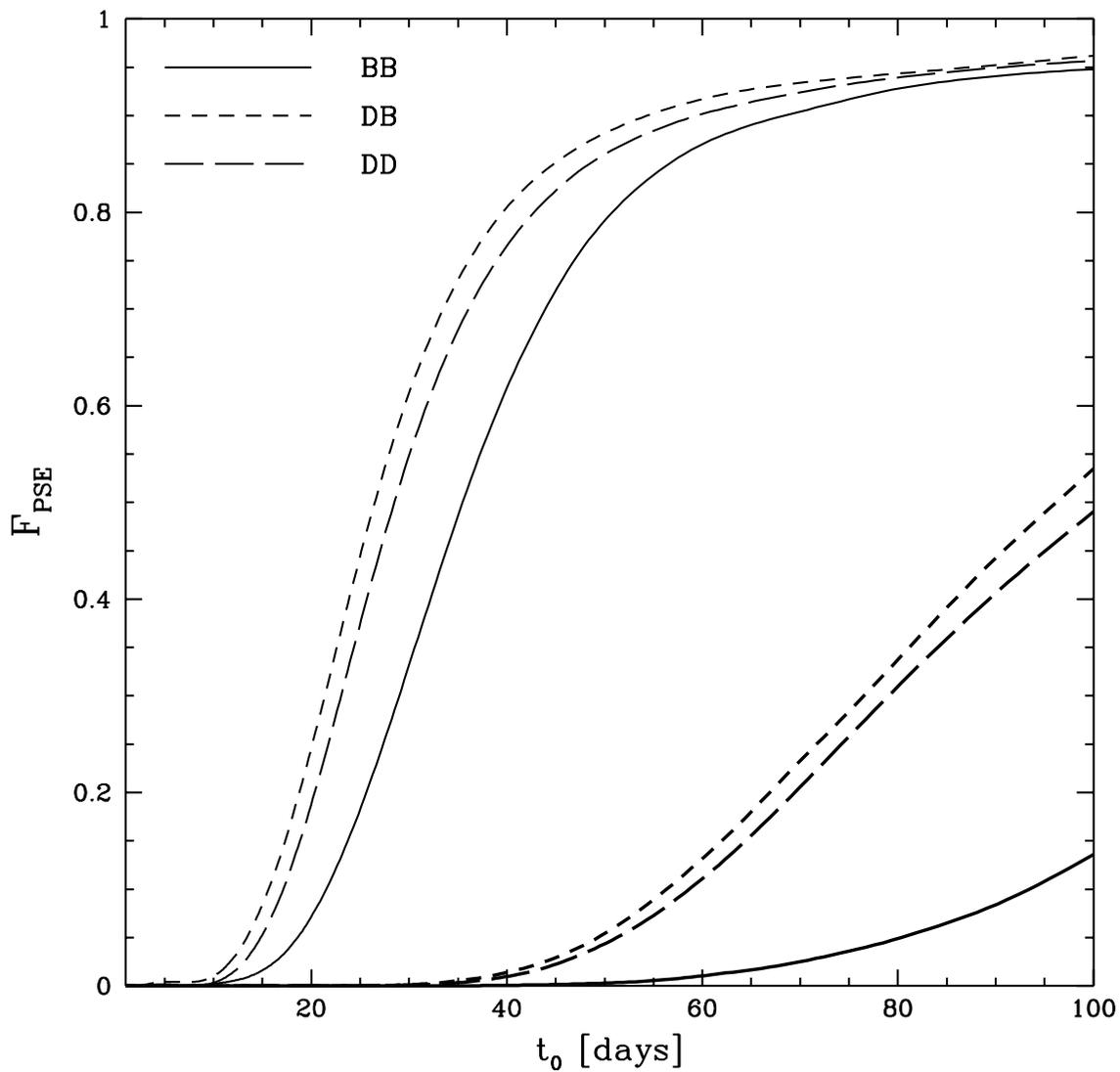}
\caption{$F_{PSE}$ as a function of $t_0$ for the BB, DD, and DD
configurations. The thin lines are the results from simulated data with
0.5-hour sampling and 0.005-mag errors, illustrating what may be achieved by
intensive followup programs. The thick lines show the results from data taken
daily with 0.05-mag errors, typical of the MACHO and OGLE experiments.}
\end{figure}

In the above calculations we assumed a logarithmic MF down to 0.1 $\Msolar$ for
disk lenses, which is consistent with current ML observations. However, other
evidence points to a disk MF which turns over at masses near 0.4 $\Msolar$
(Scalo 1986; Larson 1986; Gould, Bahcall, \& Flynn 1995). Thus, given the
sensitivity of $F_{PSE}$ to disk lenses, we also consider the HST MF,
$\log{\phi} = 1.35 - 1.34 \log(\mass/{\Msolar}) -
1.85\left[\log(\mass/{\Msolar})\right]^{2}$ in the range from $0.1\ \Msolar
\leq {\mass_l} \leq 5.0\ \Msolar$, as well as a composite power-law MF with
$dN/d\mass = 0$ for $0.1\ {\Msolar} \leq {\mass_l} \leq 0.4\
\Msolar$, $dN/d\mass \; {\propto} \; \mass^{-1.25}$ for $0.4\ {\Msolar} \leq
{\mass_l} \leq 1.0\ \Msolar$, $dN/d\mass \; {\propto} \; \mass^{-2.3}$ for
$1.0\ {\Msolar} \leq {\mass_l} \leq 3.0\ \Msolar$ and $dN/d\mass \; {\propto}
\; \mass^{-3.2}$ for $3.0\ {\Msolar} \leq {\mass_l} \leq 5.0\ \Msolar$ (Mihalas
\& Binney 1982).
The HST MF has a higher mean mass than the logarithmic MF, and the composite
power-law MF favors still higher masses. The results for these MFs are compared
with our previous results in Figure 5. Parallax effects are seen to be quite
sensitive to the disk MF, becoming more dramatic with higher mean masses, and
like color shifts, may be used as a means to discriminate between different
forms for the stellar MF.
\begin{figure}
\plotone{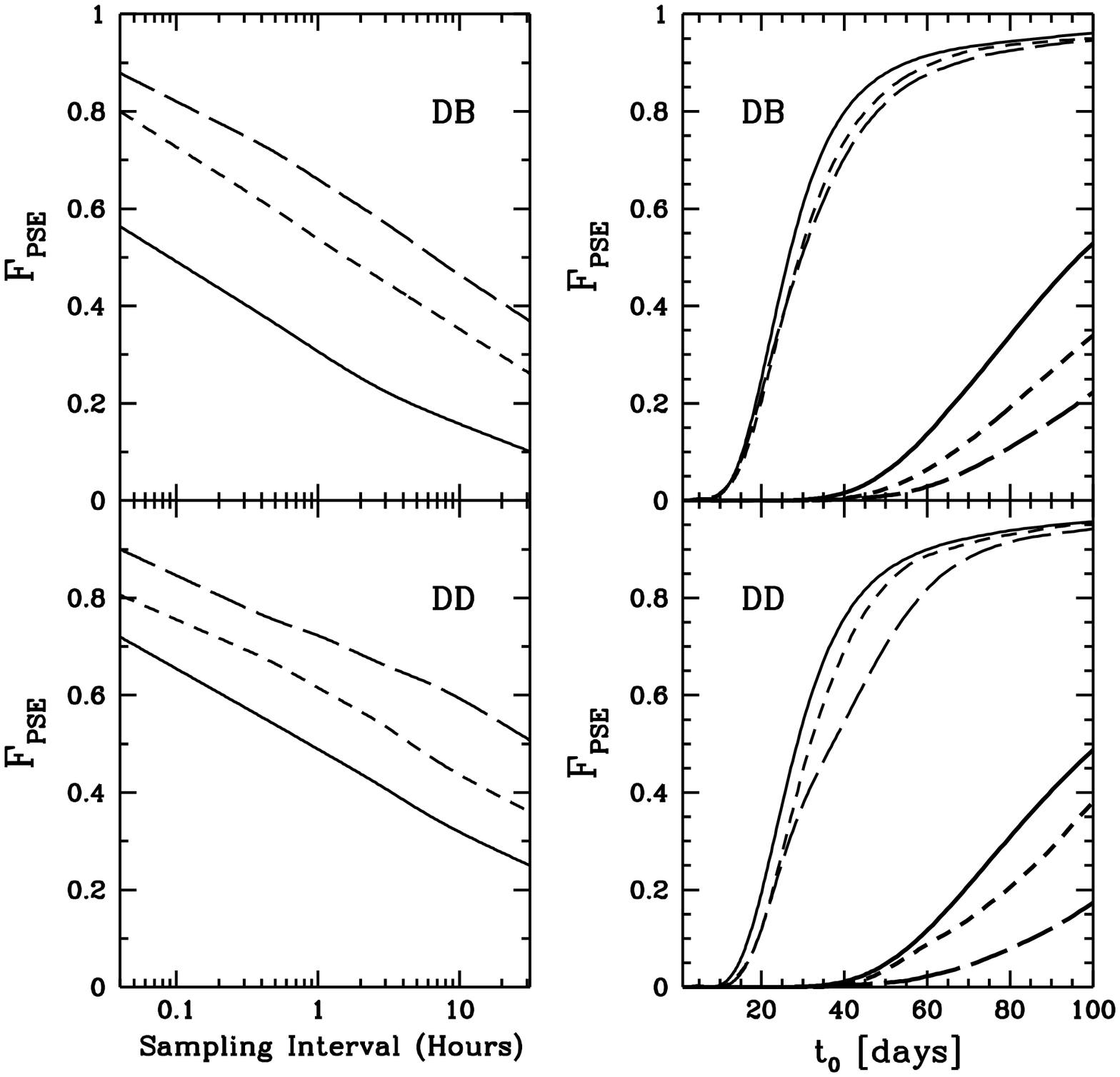}
\caption{Variations in $F_{PSE}$ arising from different disk MFs. The solid,
short-dashed, and long-dashed curves correspond to the logarithmic, HST, and
composite power-law MFs, respectively. The left panels depict the expected
fraction of PSEs as a function of sampling interval, assuming 0.01-mag errors.
The panels on the right show $F_{PSE}$ as a function of $t_0$, where the thin
lines correspond to 0.5-hour sampling with 0.005-mag errors, and the thick
lines correspond to daily sampling with 0.05-mag errors. Note that while the
overall fractions increase with higher mean lens masses, the fraction as a
function of $t_0$ are lower because the higher masses give rise to longer event
timescales.}
\end{figure}

In the preceeding discussion, we have not focused on the results from a
particular survey, but rather have considered a broad spectrum of observational
capabilities, some of which are outside the range of current surveys. For
example, 15-minute sampling with 1\% $V$-band photometry of a $V=21$ star in
Baade's Window during full moon with 1'' seeing would require a 3-meter
telescope for the followup surveys. Photometric errors will generally depend on
the source brightness, and the details of any given experiment are ultimately
governed by systematics, such as seeing variations, varying aberrations, and
telescope alignment errors, that will require more detailed modelling. Our
results for the fraction of observably distorted events arise from the
parameterization of an idealized survey, and are intended as a template for
comparison with various observational assumptions. For example, if the above
systematic errors give rise to night-to-night variations in the photometry
which are of order 0.01 mag (larger than
the assumed statistical errors), a lower limit to $F_{PSE}$ could be attained
by assuming a degraded photometric error of 0.01 mag (the magnitude of the
nightly errors) and a degraded sampling frequency of one per day. In this case
Figure 3 predicts fractions of 2\% and 11\% for BB and DB lensing,
respectively, for an overall rate of about 4\%. It should be noted, however,
that dedicated ground-based instrumentation taking continuous and redundant
measurements from multiple sites, together with possible space-based
observations, should help control such effects. Moreover, since it is the
source amplification, and not absolute brightness, which is essential,
calibration against non-variable sources in the survey may also help reduce the
effects of long-term drifts. As mentioned above, techniques involving relative
photometry are capable of yielding high precision measurements, particularly
if coupled with space-based observations to provide a stable reference frame.

Depending on the particular capabilities of the microlensing followup surveys,
10\% or more of all events observed toward the bulge are expected to have a
detectable parallax shift at the 95\% confidence level. Such events can yield
valuable information about the properties of the lens (Alcock et al.~1995b). In
particular, since
\begin{equation}
\alpha = \frac{R_{\oplus}}{R_e}(1 - x) \;= \frac{\omega{R_{\oplus}}}{v}(1-x),
\label{alpha2}
\end{equation}
measurements of $\alpha$ and $\omega$ allow the determination of the reduced
transverse speed $\tilde{v} = v/(1-x) = \omega{R_{\oplus}}/{\alpha}$, giving
$v$ as a function of $x$ (or $D_{ol}$, if $D_{os}$ is known). In our
simulation, analysis of the covariance matrix of the 6-parameter fit (see BKR)
yields the standard errors in $t_0 = {\omega}^{-1}$ and $\alpha$ for a given
event, thus determining the error in $\tilde{v}$. As a simple illustration of
the accuracies which may be attained, we simulated the light curves of events
with typical BB and DB configurations, i.e., a 0.3-$\Msolar$ lens at 6 kpc (the
near end of the bulge) and 4 kpc, respectively, lensing a source at 8 kpc, with
$u_{\min} = 0.2$, $\phi = 1$ rad, and $t_{\max}$ at June 1. The resulting
$1\sigma$ limits on $\tilde{v}$, for various timescales and observational
parameters, are shown in Table 1. While the best
results are obtained for the longest-duration events, useful constraints can be
obtained for shorter duration events ($t_0 \approx 1$ month) as well.
\begin{table}
\begin{center}
\begin{tabular}{|c|c|c||c|c|} \hline
\multicolumn{5}{|c|}{$\Delta{\tilde{v}}/\tilde{v}$~~~($1\sigma$ limits)}  \\
\hline
\multicolumn{3}{|c||}{ } & 1 hr., 0.01 mag & 0.5 hr., 0.005 mag  \\ \hline
\hline
\multicolumn{1}{|c|}{ } & & 15 & 2.25 & 0.79 \\ \cline{2-5}
BB & $t_0$ (days) & 30 & 0.29 & 0.10 \\ \cline{2-5}
& & 60 & 0.04 & 0.01 \\ \hline \hline
& & 15 & 1.33 & 0.47 \\ \cline{2-5}
DB & $t_0$ (days) & 30 & 0.17 & 0.06 \\ \cline{2-5}
& & 60 & 0.03 & 0.01 \\ \cline{2-5}
& & 100 & 0.01 & 0.004 \\ \hline
\end{tabular}
\end{center}
\caption{Estimated $1\sigma$ accuracies for measurements of $\tilde{v}$ from
the simulated BB and DB events described above, for various event timescales,
$t_0$. Shown are the results from simulated data taken with hourly sampling and
0.01-mag errors, and from 0.5-hourly sampling and 0.005-mag errors. Note that
useful accuracies are achieved even for events with $t_0 \approx 30$ days.}
\end{table}

Gaudi \& Gould (1996) have performed a similar calculation for PSEs observed
simultaneously from the Earth and by a satellite in heliocentric orbit and find
that for photometric precisions of 1\% to 2\%, such observations could measure
$\tilde{v}$ to an accuracy of $\leq 10$\%, at the $1{\sigma}$ level, for over
70\% of disk lenses and over 60\% of bulge lenses. Adopting their $u_{\min}$
distribution, we find that for hourly sampling and 1\% photometric precision
(typical of the PLANET data), ground-based observations of the parallax effect
will measure $\tilde{v}$ to this accuracy in only 2\% of all BB events and 14\%
of all DB events, indicating that satellite measurements would provide
considerable improvement.

Parallax-shift analysis provides a constraint between $v$ and $x$ which removes
one degree of degeneracy from $t_0$; knowledge of $\tilde{v}$ also allows one
to express $\mass_l$ in terms of $x$. In particular,
\begin{equation}
t_0 = \frac{R_e}{v} = \frac{\sqrt{4G{\mass_l}x(1-x)D_{os}/c^2}}{\tilde{v}(1-x)}
\, \Longrightarrow \, \mass_{l}(x) =
\frac{\tilde{v}^{2}{t_0}^{2}c^{2}}{16GD_{os}} \frac {1-x}{x}.
\end{equation}
Moreover, knowledge of $\phi$ provides additional information about the
direction of the relative transverse velocity which can be useful in
distinguishing between lenses belonging to bulge and disk populations. Such
constraints can be used in conjunction with those obtained from other
independent analyses, such as deviations from the point-source approximation,
or analysis of blending effects.

\section{CORRECTIONS TO MICROLENSING DISTRIBUTIONS}

We have already seen that if a shape distorted event is fit with the standard
3-parameter curve, the inferred values of $u_{\min}$, $t_{\max}$, and $t_0$ can
be appreciably in error (Alcock et al.~1995b; BKR). While the first two are
essentially random variables and do not affect our understanding of the lens
distributions, the event timescale, $t_0$, is critical, and any discrepancy in
this value will propagate into the assumed values for the lens distributions
and the overall optical depth. In principle, every ML event is parallax shifted
and (if the
lenses are not assumed to be totally dark) color shifted; even if the lens is
dark, the observed light curve may be distorted by blended light from an
optical companion (Di Stefano \& Esin 1995). While such effects are important
in that they allow us to
derive more information from an event than is otherwise possible, and both
effects may, in some fraction of events, be highly pronounced, in practice we
find that their
magnitude is usually small. In the case of blending, this is simply because one
is usually searching for the small contribution of a faint star (possibly the
lens) to the amplified light of a brighter star. For the parallax effect, the
typical mass range of the lenses, together with the microlensing geometry
toward the bulge, conspire to make $\alpha$ typically small. The degree to
which these effects are observed is primarily a function of the resolution of
the data, and many of the current surveys are not capable of detecting them.
Thus, we wish to examine whether any systematic errors are incurred by
routinely applying standard 3-parameter fits to ML light curves which are
allowed to suffer both parallax and blending distortions. To do this, we once
again employ a Monte Carlo simulation of lensing events, using the Galactic
model described in BKR. The shape distortion in a given event (parameterized by
$\alpha$ and the luminosity offset ratio $r$) is determined by the masses and
distances of the source and lens; the parallax effect is modelled as
described above, and for simplicity,
blending is assumed due to unlensed light from the lens, following the model
of
BKR. This assumption reflects an upper limit to the impact of blending, since
the lens is located nearer to us than an optical companion is likely to be,
thus appearing brighter, and it assumes blending occurs in every event. Each
such light curve is generated and then fit for only the three standard
parameters to yield $t_{0}^{\inf}$, $t_{\max}^{\inf}$, and $u_{\min}^{\inf}$,
where the superscripts distinguish these inferred values from the actual values
(no superscript) used to generate the events. Due to the presence of $\alpha$
and $r$, the inferred values will generally differ from the actual ones so that
the inferred distributions of impact parameters (i.e., peak amplifications) and
timescales, will be different than their actual values. The model we employ is
intended simply to give an estimate of the expected magnitude of such
deviations.

Before turning to the calculations, it is instructive to make some brief
qualitative observations about the expected differences between
the inferred and actual quantities. The parallax effect is essentially a
low-amplitude, sinusoidal-type variation superimposed on $u(t)$. On average, it
should not push either the peak amplification (given by $u_{\min}$) or the
event timescale, $t_0$, in any particular direction, since the correction to
$u(t)$ in Eq.~(\ref{u(t)}) can be positive or negative. Blending distortions
(arising from the lens or otherwise), however, do operate preferentially.
Suppose the light from an unresolved star (possibly the lens) with apparent
brightness $\ell_*$ is blended with that of a source with a baseline brightness
$\ell_s$. If the amplification of the background star is $A$, then the observed
brightness is $\ell_{obs} = \ell_{*} + A
\ell_{s}$; the baseline brightnesses are obtained by setting $A=1$.
Due to the contribution of the unlensed light, the {\it observed}
amplification,
which we denote by $\Ampl$, is different from the microlensing
amplification $A$.  The observed amplification in a given waveband will be
\begin{equation}
     \Ampl (t)= {\ell_{*} + A(t) \ell_{s}
                    \over \ell_{*} + \ell_{s}}
                = (1-r) + A(t) r,
\label{observedamplification}
\end{equation}
where $r=\ell_{s} / (\ell_{s} +
\ell_{*})$ is the luminosity offset ratio. The presence of $r$ distorts the
shape of the observed light curve, even if the two stars have the same color.
In the case of no blending, $r=1$, and we recover the standard ML scenario. If
there is {\em any} blended light, however, $r$ decreases. It can easily be
shown that the observed amplification $\Ampl$ is a monotonically decreasing
function of $r$, so that all blended events exhibit an amplification which is
everywhere less than that which would have been observed in the absence of
blending. It follows that $u_{\min}^{\inf} > u_{\min}$ for all blended events.
Since an event is only registered when the {\em observed} amplification exceeds
threshold, the
reduction in amplification translates to a shorter event duration, and thus a
smaller timescale, $t_0$. In particular, an event will only be registered while
$\Ampl > 1.34 \Longrightarrow A \geq A_{\rm thresh}=[1.34 - (1 - r)]/r$, or
only when the dimensionless lens--line-of-sight
distance is $u \leq u_{\rm thresh} = u(A_\thresh)$, where $u_{\rm thresh} < 1$.
The observed event duration, $t_e = 2{t_0}[ u_\thresh^2 -
u_{\rm min}^2 ]^{1/2}$, is thus shorter than the no-blending case
(where $u_{\rm thresh} = 1$) leading to a decrease in the inferred $t_0$.
Thus, from blending effects, we expect to see the inferred duration
distribution shifted toward shorter timescales.
Moreover, since the blended stars are presumably low-mass dwarfs which emit
most of their light at redder wavelengths, $r$ will decrease with increasing
wavelength so the reduction in amplification and in event timescale should, in
principle, be more pronounced for longer wavelengths.

In Figure 6, we plot the smoothed, non-normalized duration distribution,
$d{\Gamma}/d{t_0}$, for BB and DB events, where we have again adopted the OGLE
efficiency, ${\epsilon}(t_0)$. The solid curves are the actual values and the
dashed curves are the results from the inferred timescales from simulated
distorted events with main-sequence (MS) source stars observed in $V$, allowed
to suffer both parallax and blending effects. All curves assume a logarithmic
disk MF. For BB events, there is no appreciable difference between the inferred
and actual distributions. This is essentially due to the shape of the bulge MF,
which favors very low-mass lenses, and the $V = 21$ magnitude cutoff for
resolved sources. As a result, only the brightest MS bulge stars (near our
upper mass cutoff) are lensed in our model, and the distortions from the
distant, low-mass blended stars are too small to systematically alter the
timescales. The same result would obviously hold true in the case of giant
sources. For DB events, however, the lenses are closer to the solar circle, and
their light contributions can become significant. Thus, for the DB curve, we
find that for $t_0 > 14$ days, the distribution is shifted by roughly 10\%
along the entire curve towards shorter $t_0$. Our calculations also show that
for the flatter disk MFs, which favor slightly higher masses, the change
becomes more dramatic ($\approx$20\% for the HST MF, and $\approx$40\% for the
composite power-law MF), affecting even events with giant sources. DD events
(not shown) show a similar behavior, though to a lesser degree than DB events,
since the sources are so much nearer as well. We also ran a simulation in which
events were distorted only by parallax, and found no significant difference
between the observed and actual distributions, confirming that the shift seen
in the Figure is due to blending.

In addition to altering the duration distribution, the errors in $t_0$ also
propagate into the optical depth, $\tau$, tending to {\em reduce} the inferred
value of $\tau$ as well. However, an exact determination of both
$d{\Gamma}/d{t_0}$ and $\tau$ requires an additional correction to account for
the discrepancy between the assumed and actual values of the source number
density. Since a given blended source may actually be two or more unresolved
sources, the inferred optical depth will as a result be {\em enhanced} relative
to the actual value, and $d{\Gamma}/d{t_0}$ may be affected as well, though it
is unlikely that events would be made longer. A precise calculation of blending
effects requires a detailed model employing the frequency and separation
distributions of optical binaries along the line of sight, and is beyond our
present scope. The above results are meant only to
illustrate the impact of some of these effects.
\begin{figure}
\plotone{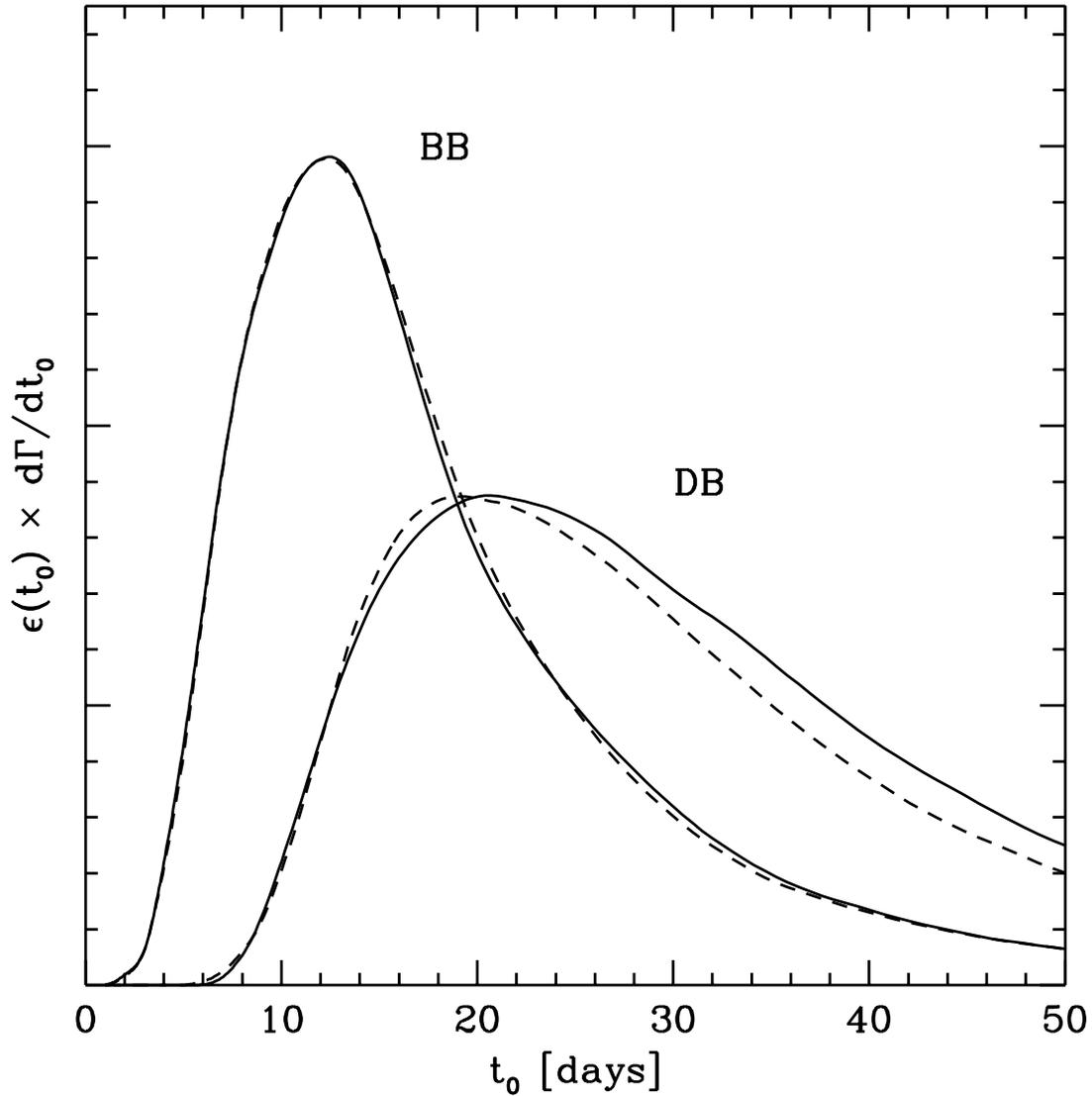}
\caption{Non-normalized microlensing event duration distributions for the bulge
and disk models used. The solid lines are the results obtained using the actual
event timescales $t_0$, while the dashed lines show the results obtained using
$t_{0}^{\inf}$ from a standard fit to a blended event. The BB distribution
(which matches the peak of the combined MACHO and OGLE data) is unaffected, but
the DB distribution is shifted by about 10\% towards smaller $t_0$.}
\end{figure}

Since $u_{\min}$, the impact parameter in units of $R_e$, is a random variable,
the distribution of $u_{\min}$ from an ensemble of events (expected to be
uniform from 0 to 1) is used as a statistical check on microlensing. Due to
blending effects, however, the distribution of $u_{\min}^{\inf}$ will not be
uniform. First, since blending always causes a reduction in the observed
amplification, the inferred impact parameter will be greater than its actual
value in every blended event, resulting in fewer $u_{\min}^{\inf}$ at low
values. However, there will also be a decrease at high values, because blended
events where $u_{\min} > u_{\rm thresh}$ (recall $u_{\rm thresh} < 1$ in
blended
events) will never register. These effects are seen in Figure 7, which shows
the non-normalized inferred and actual $u_{\min}$ number distributions for both
BB and DB events. We again assume observations in $V$ and a logarithmic disk
MF. Detection efficiencies have not been included, though it is expected that
experimental efficiencies efficiencies
should drop off at high $u_{\min}$. While the predicted behavior of $u_{\min}$
is demonstrated by the Figure, resolving this effect in practice would require
many more events than currently observed.
\begin{figure}
\plotone{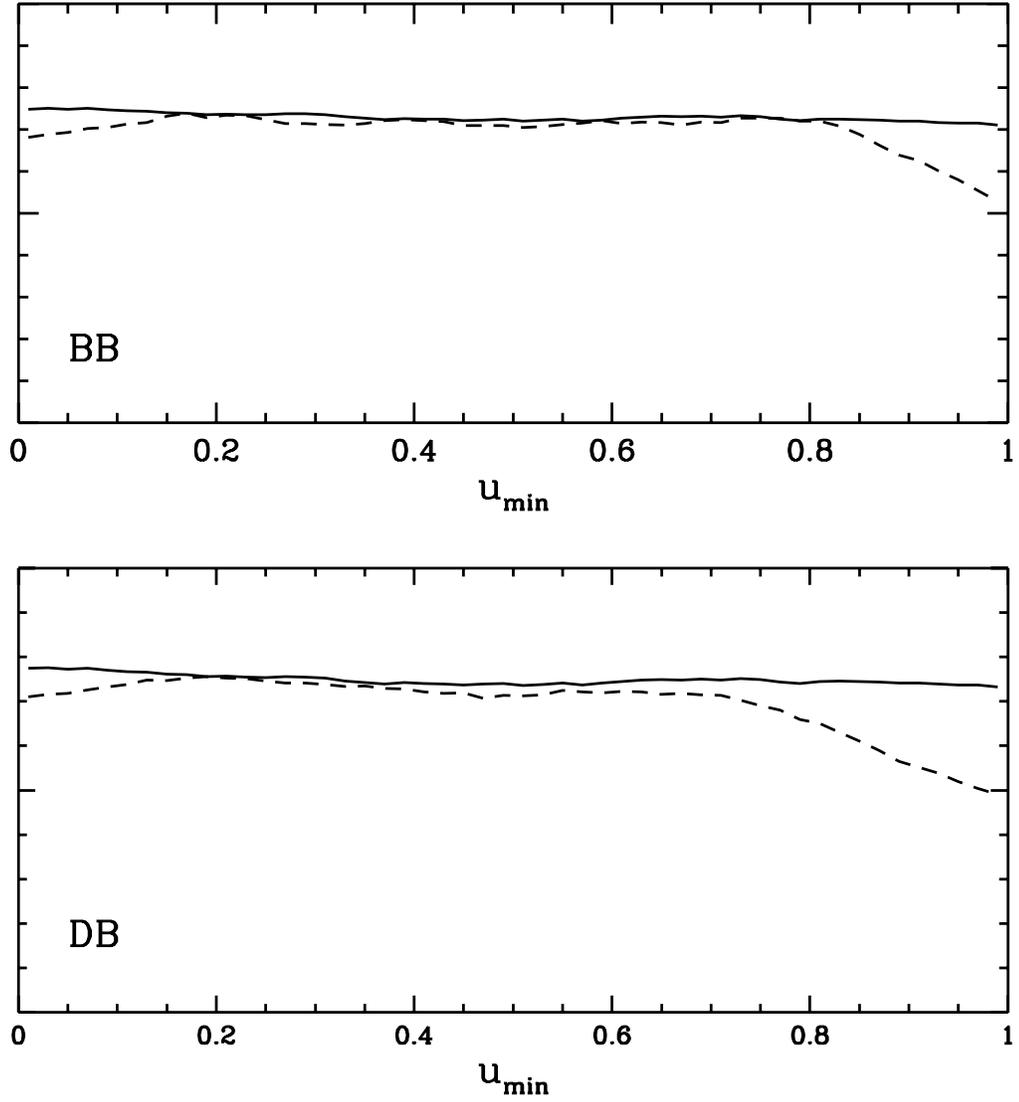}
\caption{Non-normalized number histogram of the impact parameter distribution
for BB events (upper panel) and DB events (lower panel) observed in $V$. The
solid lines show the (approximately uniform) distributions of the actual values
($u_{\min}$), while the dashed lines show the distributions of the inferred
values ($u_{\min}^{\inf}$). The latter are non-uniform due to blending
effects.}
\end{figure}

\section{CONCLUSION}

While many prospects for learning more about ML events involve intensive
followup observations aimed at resolving distortions to the standard ML light
curve, few detailed calculations of rates and features of such distortions
exist for realistic Galactic models. We have employed a detailed model to
calculate the fraction of ML events which will exhibit a significant parallax
distortion, as seen by the followup monitoring programs. We find that with
frequent and precise observations of events in progress, from 10\% up to 39\%
of all events arising from various lens-source pairings are expected to exhibit
a parallax shift, including some events with $t_0 \lesssim 2$ months. Such
events
can be used to place an additional constraint among the parameters of interest
(namely $\mass_l,D_{ol},\mbox{ and }v$) with reasonable precision, allowing one
to express $\mass_l$ as a function of $D_{ol}$ and thus better determine the
characteristics of the lensing population. Failing to account for ML
distortions, particularly blending, can lead to errors in the inferred
parameters (most notably $t_0$) which propagate into the inferred duration
distribution and impact-parameter distribution and impact the assumed overall
optical depth.

The most important application of analyzing such distortions is in discerning
whether observed bulge events are due to structure in the bulge, or excess mass
in the form of sub-stellar objects in the disk. However, the significance of
such work extends to many areas. Determining the origin of the excess events
will lead to a better understanding of Galactic structure and stellar
populations, particularly the MF of low-mass stars in the disk and bulge. The
latter relates directly to our understanding of the process of star formation.
Moreover, if the bulge of our Galaxy is representative of relaxed stellar
systems such as elliptical galaxies, then knowledge of the bulge luminosity
function and dynamical structure will have implications for galaxy formation
and evolution. Precise knowledge of the mass of the disk and bulge also
constrains the halo mass and places strong limits on the dark-matter content.
If it can be determined that there is more mass in the bulge and/or disk, this
implies less halo dark matter and has important consequences for the predicted
event rates in direct and
indirect searches for exotic dark matter (Jungman et al.~1996).

\acknowledgements

We wish to thank H. S. Zhao for providing us with data from his self-consistent
model of
the Galactic bar, and for helpful discussions. We also thank R. Olling for many
insightful suggestions on
constructing a reasonable model for the Galactic disk
as well as numerous helpful suggestions. We
thank A. Gould, E. Wright, and R. M. Rich for very useful and detailed
comments.  This work was
supported in part by the U. S. Department of
Energy under contract DE-FG02-92ER40699, and by NASA grant NAG5-3091.

\end{document}